\documentclass[runningheads]{llncs}
\usepackage{graphicx}
\usepackage{epsfig,epsf}
\usepackage{epstopdf}
\usepackage{graphics}
\usepackage{array}
\usepackage{amsmath}
\usepackage{amssymb}
\usepackage{comment}
\usepackage{algorithm}
\usepackage{algorithmic}

\usepackage{fullpage}
\date{}

\newtheorem{dfn}{Definition}
\title{FO and MSO approach to Some Graph Problems:\\ Approximation and Poly time Results}
\author{Kona Harshita ${^1}$ \and Sounaka Mishra ${^2}$ \and Renjith. P ${^1}$ \and N. Sadagopan ${^1}$} 
\institute{${^1}$ Indian Institute of Information Technology Design and Manufacturing, Kancheepuram, Chennai. \\ ${^2}$ Department of Mathematics, Indian Institute of Technology Madras, Chennai, India.\\
\email{sounak@iitm.ac.in,\{sadagopan,coe14b016,coe14d002\}@iiitdm.ac.in}}
\begin{document}
\maketitle
\begin{abstract}
The focus of this paper is two fold.  Firstly, we present a logical approach to graph modification problems such as minimum node deletion, edge deletion, edge augmentation problems by expressing them as an expression in first order (FO) logic.  As a consequence, it follows that these problems have constant factor polynomial-time approximation algorithms.  In particular, node deletion/edge deletion on a graph $G$ whose resultant is cograph, split, threshold, comparable, interval and permutation are $O(1)$ approximable.  Secondly, we present a monadic second order (MSO) logic to minimum graph modification problems, minimum dominating set problem and minimum coloring problem and their variants.  As a consequence, it follows that these problems have linear-time algorithms on bounded tree-width graphs.  In particular, we show the existance of linear-time algorithms on bounded tree-width graphs for star coloring, cd-coloring, rainbow coloring, equitable coloring, total dominating set, connected dominating set.  In a nut shell, this paper presents a unified framework and an algorithmic scheme through logical expressions for some graph problems through FO and MSO.
\end{abstract}
\section{Introduction}
Many interesting problems that arise in the field of mathematics and computing can be casted as graph-theoretic problems.  Popular ones are graph modification, subset and permutation problems.  Graph modification problems include node deletion, edge deletion, and edge augmentation problems to satisfy a given graph property.  For node (edge) deletion problems; given a connected graph $G$ and a non-trivial hereditary property $\Psi$, it is asked for a minimum set $S \subset V(G)$ ($E' \subset E(G)$ for edge deletion) such that $G\setminus S$ ($G \setminus E'$) satisfies the property $\Psi$. We shall study properties $\Psi$ such as {\em cograph, split, threshold, comparable, interval and permutation}.  Interestingly, these properties are non-trival and hereditary properties. A graph property $\Psi$ is called a non-trivial if there are infinite number of graphs satisfy $\Psi$ and infinite number of graphs do not satisfy $\Psi$. The property $\Psi$ is hereditary if $G$ satisfies $\Psi$ then all its vertex induced subgraphs satisfy $\Psi$. In some sense, for a connected graph $G$, node/edge deletion problem finds a maximal subgraph that satisfies some non-trivial property \cite{maxsubgraph}.  Interestingly, these problems have applications in several areas, such as molecular biology, networks reliability and numerical algebra \cite{appln}.  

On the complexity perspective, node/edge deletion problems with respect to non-trivial hereditary properties are NP-complete \cite{appln}. To cope up with NP-completeness, the study of polynomial-time approximation algorithms is one of the natural directions for further research on these problems.  The reduction presented by Lewis and Yannakakis \cite{appln} is generic in nature as it gives hardness result of many node-deletion problems.  Moreover, their reduction is approximation preserving and thus, any good approximation algorithm for node-deletion problem will inturn give a good approximation for minimum vertex cover problem which was the candidate NP-hard problem used in the reduction.   In an attempt to design good approximation algorithms for node-deletion problems, in \cite{appln}, a framework was presented and fine tuning of the framework is necessary to get an approximation algorithm for a specific node-deletion problem.  It is natural to ask: Is there a different framework that yields approximation algorithms for many node-deletion problems.  

Graphs can be seen in three different ways; (i) graphs as a structure (ii) graphs as sets (for example: logic) (iii) graphs as matrices (for example: spectral graph theory).   Fagin \cite{fagin} looked at graphs as sets and  initiated the {\em logical approach} to graph optimization problems.  This was a celebrated result as it made a paradigm shift in complexity perspective on how optimization problems can be looked at through a different lens.  This line of research was well appreciated by researchers Papadimitriou and Yannakakis \cite{papad}, Kolaitis and Thakur \cite{kolaitis} as they looked at graphs as sets and deepened the study of logic in the context of approximation algorithms.   In \cite{papad}, it was established that every optimization problem in MAX NP (class of maximization problems defined using existential second-order formulae) is constant-factor approximable.   Further, Kolaitis and Thakur \cite{kolaitis} showed that, if a non-trivial hereditary property $\Psi$ associated with a minimization problem can be expressed in a first order (FO) formula, then it belongs to the class MIN $F^{+} \Pi_1$.  More precisely,  minimization problem belongs to MIN $F~\Pi_1$ if it can be expressed as a {\em universal} first order formula and it belongs to  $MIN F^{+} \Pi_1$ if all the occurrences of the set $S$ (solution set) are positive.  An important contribution by \cite{kolaitis} is that every problem in MIN $F^{+}\Pi_1$ is constant-factor approximable. 

In the first part of the paper, we shall present first order logical formule for node/edge deletion problems for hereditary properties cograph, split, threshold, comparable, interval and permutation.   It follows from \cite{kolaitis} that these problems are $O(1)$ approximable.  Second order logic and in particular, monadic second order (MSO) logic is an extension of FO which is like FO give set theoretic perspective of the problem under study.  Interestingly, MSO expression helps us to comment time complexity of NP-complete problems for restricted graph classes.  While FO investigate NP-complete problems from approximation perspective, MSO investigate NP-complete problems by restricting the input. 

First order logic allows quantification of only variables that range over the elements of a set but not the set itself whereas second order logic allows quantification over sets, relations, functions etc.  Monadic second order logic is the fragment of second order logic where the second-order quantification is limited to quantification over sets.  Courcelle \cite{courcelle} showed that any graph problem expressible by monadic second order logic can be solved in linear time for bounded treewidth graphs.  This result assumes that the underlying tree decomposition of the graph is given as part of the input.  

In the second part of the paper, we shall present MSO for dominating set and its variants (subset problems) and coloring and its variants (permutation problems).  As a consequence it follows that these problems have linear-time algorithms on bounded treewidth graphs.  It is important to highlight that FO presented in this paper can be fine tuned to get MSO for node/edge deletion problems.  We thus, focus our MSO study on other classical problems which find applications in reliable computer networks and scheduling problems \cite{domappln}.  We shall present MSO for connected dominating set, total dominating set, total outer connected dominating set, star coloring, rainbow coloring, cd-coloring and equitable coloring.  To the best of our knowledge, we believe that logical approach to these problems have not been addressed in the literature and the scheme presented here can be extended to other related problems.  

\noindent
{\bf Road map:} We shall present the first order formuale for node/edge problems by considering popular hereditary properties in Section \ref{fo}.  In Section \ref{mso}, we shall discuss MSO for dominating set, coloring and its variants.  

\section{Preliminaries}
We shall present graph-theoretic preliminaries first, followed by, definitions and notations related to FO and MSO. 
\subsection{Graph Theory Preliminaries}
Throughout this paper, we follow the notations from \cite{golumbic} and \cite{west}. Let $G = (V, E)$ be a finite undirected graph where $V(G)$ is the set of vertices and $E(G)\subseteq \{\{u,v\}|u,v\in V(G),u\neq v\}$. Let $\overline{G}$ denote the complement of $G$.  Throughout this paper, we work with simple connected graphs.  A connected graph is \emph{$P_4$-free} if it does not contain a path on four vertices as an induced subgraph. A connected graph is \emph{$2K_2$-free} if it does not contain a pair of independent edges as an induced subgraph. A connected graph is \emph{$C_k$-free} if it does not contain a cycle on $k$ vertices as an induced subgraph. 

Next, we mention a few characterizations for graphs satisfying some of the properties mentioned earlier and their proofs can be obtained from \cite{andras}.
A graph $G$ is a \emph{cograph} if every induced subgraph $H$ of $G$ with at least two vertices is either disconnected or the complement of a disconnected graph. A graph is a \emph{cograph} if and only if it is \emph{$P_4$-free}.  A graph $G$ is a \emph{split graph} if the vertices can be partitioned into an independent set and a clique.  A graph is a \emph{split graph} if and only if it is \emph{$\{2K_2,C_4,C_5\}$-free}. A graph $G$ is a \emph{threshold graph} if and only if it is \emph{$\{2K_2,C_4,P_4\}$-free}.  A graph is a \emph{comparability graph} if it has a \emph{transitive orientation} of its edges, that is, an orientation $F$ for which $\overrightarrow{(a,b)}, \overrightarrow{(b,c)}\in F$ implies $\overrightarrow{(a,c)}\in F$. A graph is an \emph{interval graph} if its vertices can be assigned to intervals on the real line such that there is an edge between two vertices if and only if their corresponding intervals intersect. This set of intervals is called an interval model for the graph. A graph $G$ is said to be an \emph{interval graph} if and only if it is \emph{$C_4$-free} and $\overline{G}$ has a \emph{transitive orientation}. A graph $G$ is a \emph{permutation graph} if and only if both $G$ and $\overline{G}$ are \emph{comparability graphs}. 

\subsection{FO Preliminaries}
For logic, we follow the notations given in \cite{kolaitis,courcelle}. \emph{First order formulas} for a specific graph-theoretic problem is constructed using atomic formulae and logical connectives.  The variables used in the expression are elements of $V(G)$ or $E(G)$ which are called {\em universe of discourse}.  For example, the atomic formula $E(x,y)$ expresses the adjacency between $x \in V(G)$ and $y \in V(G)$, and the formula $x = y$ expresses the equality between $x$ and $y$.  We make compound logical expressions using logical connectives such as $\neg$ (negation), $\land$ (conjunction), 
$\lor$ (disjunction), $\rightarrow$ (implication) and $\leftrightarrow$ (bi-implication).   Quantification of members of the vertex set is done using two quantifiers, namely, existential quantification ($\exists x$) and universal quantification ($\forall x$).   For example, to say a graph $G=(V,E)$ is an undirected and loop-free, we write in first-order formula $\varphi$ as 
\[\varphi \equiv (\forall x)(\forall y)[(E(x,y) \leftrightarrow E(y,x)) \land x\neq y].\]
For a graph $G$ and any minimization problem to be solved on $G$, let $opt(G)$ denote the size of any optimum solution which is defined as 
\[opt_{\varrho}(G) = min_{S}\{|S|:G \models \phi(y,S)\}.\]
where $\phi(y,S)$ denotes a first-order formula; $y$ denotes the set $\{y_1,\ldots,y_k\}$ of variables used in the first-order formula and $S$ denotes the solution set. 

A formula written in first order logic is equivalent to a formula in \emph{prenex normal form} if all quantifiers are written to the left of all the other symbols. Let $\Pi _{n}$ be the class of first-order formula in prenex normal form with $n-1$ alternations of quantifiers, starting with a block of universal quantifiers. That is, formula having only a block of universal quantifiers belong to $\Pi _1$, formula having a block of universal quantifiers followed by a block of existential quantifiers belong to $\Pi _2$ and so on. The class of quantifier-free formula is denoted by $\Pi _0$.

\begin{dfn} {\rm \cite{kolaitis}}
\label{def1}
Let MIN F$^{+}\Pi _1$ (F stands for \emph{feasible}), be the class of all minimization problems that are definable using $\Pi _1$ (universal) formulae that are positive in $S$. In other words, MIN F$^{+}\Pi _1$ is the collection of all minimization problems $\varrho$ whose optimum can be expressed as
\[opt _{\varrho}(G) = min _{S}\{|S|:(G, S) \models (\forall y)\psi (y, S)\}\] 
where $S$ is a single predicate and $\psi (y, S)$ is a quantifier-free formula in which all occurrences of $S$ are positive.
\end{dfn}
For example, the problem MIN VERTEX COVER (MIN\_VC) belongs to the class MIN F$^{+}\Pi _1$ as the optimum solution for a graph $G=(V,E)$ is given by
\[opt _{MIN\_VC}(G) = min _{S}\{|S|:(G, S) \models (\forall x)(\forall y)[E(x, y) \rightarrow (S(x) \lor S(y))]\}\]

\noindent Similarly, the problem MIN DOMINATING SET (MIN\_DOM\_SET) belongs to the class MIN F$^{+}\Pi _2$ as optimum solution for a graph $G=(V,E)$ is given by
\[opt _{MIN\_DOM.\_SET}(G) = min _{S}\{|S|:(G, S) \models (\forall x)(\exists y)(S(x) \lor (S(y) \land E(x, y)))\}\]

\begin{proposition}{\rm \cite{kolaitis}}
\label{prop2}
Every problem in MIN F$^{+}\Pi _1$ are constant-approximable.
\end{proposition}

\subsection{MSO Preliminaries}
First order formulas do not allow quantification over the sets whereas \emph{Monadic second order formulas} allow universal and existential quantification over sets. In the language of graphs, they can quantify over set of vertices and edges. Monadic second-order formulas that do not contain second-order quantifiers but have free set variables are viewed as first order formulas as free set variables are same as unary relation symbol.  For example, the following MSO sentences \emph{Chordal} and \emph{Chordal\_Bipartite} says that the graph $G=(V,E)$ is chordal and chordal bipartite respectively. 
\begin{align*}
& Chordal(G) = \neg \Big [(\exists C\subseteq V)(\forall x\in C)(\exists y\in C)(\exists z\in C)\big (y\neq z \land E(x,y) \land E(x,z) \land (\forall w\in [C-(y,z)]) \\ & (\neg E(x,w)) \land |C|>3\big )\Big ] \land connectedness(C)
\end{align*}
which ensures that $G$ contains no induced cycle of length greater than 3. 
\begin{align*}
& Chordal\_Biparite(G) = \Big [(\exists R)(\exists B)(\forall x\in V)(R(x) \lor B(x)) \land \big ((\forall x\in V)(\forall y\in V)((R(x) \land R(y))(B(x) \land B(y)) \rightarrow \\ & \neg E(x,y))\big )\Big ] \land \neg \Big [(\exists C\subseteq V)(\forall x\in C)(\exists y\in C)(\exists z\in C)\big (y\neq z \land E(x,y) \land E(x,z) \land (\forall w\in [C-(y,z)])(\neg E(x,w))\big )  \\ & \land |C|>4\Big ] \land connectedness(C)
\end{align*}
which ensures that $G$ is bipartite with two color classes $R$ and $B$ and $G$ contains no induced cycle of length greater than 4. Here,
\begin{align*}
& Connectedness(C) = (\forall x)(\forall y)(C(x) \land C(y) \land (E(x,y) \lor (\exists P\subset S \land (x_1,x_2,\ldots ,x_{k})\in P \land E(x,x_1) \land \\ & (\bigwedge_{1\leq i \leq k-1} E(x_{i},x_{i+1})) \land E(x_{k},y))))
\end{align*}

\begin{proposition} {\rm \cite{courcelle}}
Every fixed MSO property can be solved in linear time on graphs with bounded tree width.
\end{proposition}

\noindent Graph classes with bounded tree width include forests, Series parallel networks, Outerplanar graphs, Halin graphs, Pseudoforests, Cactus graphs etc. 

\section{Results on FO}
\label{fo}
%Assume now that $\Psi$ is a property of finite graphs that is definable using a universal first-order sentence. Then the node deletion problem NODE-DEL $_{\Psi}$ associated with $\Psi$ is contained in the class  MIN $F ^{+} \Pi _{1}$. 

If $\Psi$ is definable by the universal sentence $(\forall x_1) \ldots (\forall x_{t})\psi(x_1, \ldots ,x_{t})$, then the optimum of NODE-DEL $_{\Psi}$ on a graph $G$ can be expressed as 
\[opt_{NODE-DEL _{\Psi}}(G) = min _{S}\{|S|:G \models (\forall x_1) \ldots (\forall x_{t})(\neg \psi(x_1, \ldots ,x_{t}) \rightarrow (S(x_1) \lor \ldots \lor S(x_{t})))\}.\]  Therefore, the node deletion problem NODE-DEL $_{\Psi}$ associated with $\Psi$ is contained in the class  MIN $F ^{+} \Pi _{1}$.  Note that the set $\{x_1,\ldots,x_t\} \subset V(G)$ and they violate the property $\Psi$.  Similarly, if $\Psi$ is definable by the universal sentence $(\forall x_1)(\forall y_1) \ldots (\forall x_{t})(\forall y_{t})(\neg \psi(x_1,y_1, \ldots ,x_{t},y_{t}))$, then the optimum of EDGE-DEL $_{\Psi}$ on a graph $G$ can be expressed as \\
\[opt_{EDGE-DEL _{\Psi}}(G) = min _{S}\{|S|:G \models (\forall x_1)(\forall y_1) \ldots (\forall x_{t})(\forall y_{t})(\neg \psi(x_1,y_1, \ldots ,x_{t},y_{t}) \rightarrow (S(x_1,y_1) \lor \ldots \lor S(x_{t},y_{t})))\}.\]   Thus, the edge deletion problem EDGE-DEL $_{\Psi}$ associated with $\Psi$ is contained in the class  MIN $F ^{+} \Pi _{1}$.   Here, $S$ denotes the solution set and $\psi$ is a quantifier-free formula. Note that $\{x_{i}, y_{i}\}\in E(G)$, $1\leq i\leq t$ and they violate the property $\Psi$.
 
\begin{proposition}{\rm \cite{kolaitis}}
\label{prop1}
NODE\_DEL$_{\Psi}$ and EDGE\_DEL$_{\Psi}$ are \emph{constant-approximable}.
\end{proposition}

We shall now present universal FO for cograph, split, threshold, comparable, interval and permutation and thereby show that node/edge deletion problems are constant factor approximable and belongs to $O(1)$ approximation class for these properties.

Let $S(x)$ and $S(x,y)$ denote the solution set of node and edge deletion problems, respectively, and $E(x,y)$ denote the adjacency between the vertices $x$ and $y$ in the graph $G$. By using these predicates, we formally define the node/edge deletion problems as follows.

\vspace{.3cm}
\noindent 
MIN COGRAPH NODE DELETION:  \\
Input:  A graph $G = (V, E)$.\\
Output:  MIN $S\subseteq V(G) \ni G\setminus S$ is a cograph.\\[0.3cm]
The optimum solution for the problem can be expressed as follows:
\begin{align*}
& opt _{MIN\_COGRAPH\_NODE\_DEL}(G) = min _{S}\big \{|S|:(G, \text{ } S) \models (\forall x_1)(\forall x_2)(\forall x_3)(\forall x_4)\big (E(x_1, x_2) \land E(x_2, x_3) \land \\ & E(x_3, x_4) \land \neg E(x_1, x_3) \land \neg E(x_1, x_4) \land \neg E(x_2, x_4) \rightarrow S(x_1) \lor S(x_2) \lor S(x_3) \lor S(x_4)\big )\big \}.
\end{align*}
which ensures that whenever there is an induced $P_4$ on four vertices $x_1 ,x_2, x_3$ and $x_4$ in $G$, atleast one of the four vertices is included in the solution set $S$.\\

\noindent MIN COGRAPH EDGE DELETION:\\
Input:  A graph $G = (V, E)$.\\
Output:  MIN $S\subseteq E(G) \ni G=(V,E\setminus S)$ is a cograph.\\[0.3cm] 
The optimum solution for the problem can be expressed as follows:
\begin{align*}
& opt _{MIN\_COGRAPH\_EDGE\_DEL}(G) = min _{S}\big \{|S|:(G, \text{ } S) \models (\forall x_1)(\forall x_2)(\forall x_3)(\forall x_4)\big (E(x_1, x_2) \land E(x_2, x_3) \land \\ & E(x_3, x_4) \land \neg E(x_1, x_3) \land \neg E(x_1, x_4) \land \neg E(x_2, x_4) \rightarrow S(x_1, x_2) \lor S(x_2, x_3) \lor S(x_3, x_4)\big )\big \}.
\end{align*}
which ensures that whenever there is an induced $P_4$ on four vertices $x_1 ,x_2, x_3$ and $x_4$ in $G$, atleast one of the three edges in $P_4$ is included in the solution set $S$. \\

\noindent MIN SPLIT NODE DELETION: \\
Input:  A graph $G = (V, E)$.\\
Output: MIN $S\subseteq V(G) \ni G\setminus S$ is a split graph. \\[0.3cm] 
The optimum solution for the problem can be expressed as follows:
\begin{align*}
& opt _{MIN\_SPLIT\_NODE\_DEL}(G) = min _{S}\Big \{|S|:(G, S) \models \Big [(\forall x_1)(\forall x_2)(\forall x_3)(\forall x_4)\big (E(x_1, x_3) \land E(x_2, x_4) \land \\ & \neg E(x_3, x_4) \land \neg E(x_1, x_2) \land \neg E(x_1, x_4) \land \neg E(x_2, x_3) \rightarrow S(x_1) \lor S(x_2) \lor S(x_3) \lor S(x_4)\big )\Big ]\land \Big [(\forall x_1)(\forall x_2)(\forall x_3)\\ & (\forall x_4)\big (E(x_1, x_2) \land E(x_2, x_3) \land E(x_3, x_4) \land E(x_1, x_4) \land \neg E(x_1, x_3) \land \neg E(x_2, x_4) \rightarrow S(x_1) \lor S(x_2) \lor S(x_3) \lor \\ & S(x_4)\big )\Big ]\land \Big [(\forall x_1)(\forall x_2)(\forall x_3)(\forall x_4)(\forall x_5)\big (E(x_1, x_2) \land E(x_2, x_3) \land E(x_3, x_4) \land E(x_4, x_5) \land E(x_1, x_5) \land \\ & \neg E(x_1, x_3) \land \neg E(x_1, x_4) \land \neg E(x_2, x_4) \land \neg E(x_2, x_5) \land \neg E(x_3, x_5) \rightarrow S(x_1) \lor S(x_2) \lor S(x_3) \lor S(x_4) \lor S(x_5)\big )\Big ]\Big \}.
\end{align*}
which ensures that whenever there is an induced $2K_2$ or $C_4$ on four vertices $x_1 ,x_2, x_3$ and $x_4$, atleast one of the four vertices is included in the solution set $S$ and whenever there is an induced $C_5$ on five vertices $x_1 ,x_2, x_3, x_4$ and $x_5$, atleast one of the five vertices is included in the solution set $S$.\\

\noindent MIN SPLIT EDGE DELETION: \\
Input:  A graph $G = (V, E)$.\\
Output:  MIN $S\subseteq E(G) \ni G=(V,E\setminus S)$ is a split graph.\\[0.3cm] 
The optimum solution for the problem can be expressed as follows:
\begin{align*}
& opt _{MIN\_SPLIT\_EDGE\_DEL}(G) = min _{S}\Big \{|S|:(G, S) \models \Big [(\forall x_1)(\forall x_2)(\forall x_3)(\forall x_4)\big (E(x_1, x_3) \land E(x_2, x_4) \land \\ & \neg E(x_3, x_4) \land \neg E(x_1, x_2) \land \neg E(x_1, x_4) \land \neg E(x_2, x_3) \rightarrow S(x_1, x_3) \lor S(x_2, x_4)\big )\Big ] \land \Big [(\forall x_1)(\forall x_2)(\forall x_3)(\forall x_4)\\ & \big (E(x_1, x_2) \land E(x_2, x_3) \land E(x_3, x_4) \land E(x_1, x_4) \land \neg E(x_1, x_3) \land \neg E(x_2, x_4) \rightarrow S(x_1, x_2) \lor S(x_2, x_3) \lor S(x_3, x_4) \\ & \lor S(x_1, x_4)\big )\Big ] \land \Big [(\forall x_1)(\forall x_2)(\forall x_3)(\forall x_4)(\forall x_5)\big (E(x_1, x_2) \land E(x_2, x_3) \land E(x_3, x_4) \land E(x_4, x_5) \land E(x_1, x_5) \land \\ & \neg E(x_1, x_3) \land \neg E(x_1, x_4) \land \neg E(x_2, x_4) \land \neg E(x_2, x_5) \land \neg E(x_3, x_5) \rightarrow S(x_1, x_2) \lor S(x_2, x_3) \lor S(x_3, x_4) \lor \\ & S(x_4, x_5) \lor S(x_1, x_5)\big )\Big ]\Big \}.
\end{align*}
which ensures that whenever there is an induced $2K_2$ or $C_4$ on four vertices $x_1 ,x_2, x_3$ and $x_4$, atleast one of the two or four edges is included in the solution set $S$ and whenever there is an induced $C_5$ on five vertices $x_1 ,x_2, x_3, x_4$ and $x_5$, atleast one of the five edges is included in the solution set $S$.\\

\noindent MIN THRESHOLD NODE DELETION: \\
Input:  A graph $G = (V, E)$.\\
Output:  MIN $S\subseteq V(G) \ni G\setminus S$ is a threshold graph.\\[0.3cm] 
The optimum solution for the problem can be expressed as follows:
\begin{align*}
& opt _{MIN\_THRES\_NODE\_DEL}(G) = min _{S}\Big \{|S|:(G, S) \models \Big [(\forall x_1)(\forall x_2)(\forall x_3)(\forall x_4)\big (E(x_1, x_3) \land E(x_2, x_4) \land \\ & \neg E(x_3, x_4) \land \neg E(x_1, x_2) \land \neg E(x_1, x_4) \land \neg E(x_2, x_3) \rightarrow S(x_1) \lor S(x_2) \lor S(x_3) \lor S(x_4)\big )\Big ]\land \Big [(\forall x_1)(\forall x_2)(\forall x_3)\\ & (\forall x_4)\big (E(x_1, x_2) \land E(x_2, x_3) \land E(x_3, x_4) \land E(x_1, x_4) \land \neg E(x_1, x_3) \land \neg E(x_2, x_4) \rightarrow S(x_1) \lor S(x_2) \lor S(x_3) \lor \\ & S(x_4)\big )\Big ]\land \Big [(\forall x_1)(\forall x_2)(\forall x_3)(\forall x_4)\big (E(x_1, x_2) \land E(x_2, x_3) \land E(x_3, x_4) \land \neg E(x_1, x_3) \land \neg E(x_1, x_4) \land \neg E(x_2, x_4) \\ & \rightarrow S(x_1) \lor S(x_2) \lor S(x_3) \lor S(x_4)\big )\Big ]\Big \}.
\end{align*}
which ensures that whenever there is an induced $2K_2, C_4$ or $P_4$ on four vertices $x_1 ,x_2, x_3$ and $x_4$, atleast one of the four vertices is included in the solution set $S$.\\

\noindent MIN THRESHOLD EDGE DELETION: \\
Input:  A graph $G = (V, E)$.\\
Output:  MIN $S\subseteq E(G) \ni G=(V,E\setminus S)$ is a threshold graph.\\[0.3cm] 
The optimum solution for the problem can be expressed as follows:
\begin{align*}
& opt _{MIN\_THRES\_EDGE\_DEL}(G) = min _{S}\Big \{|S|:(G, S) \models \Big [(\forall x_1)(\forall x_2)(\forall x_3)(\forall x_4)\big (E(x_1, x_3) \land E(x_2, x_4) \land \\ & \neg E(x_3, x_4) \land \neg E(x_1, x_2) \land \neg E(x_1, x_4) \land \neg E(x_2, x_3) \rightarrow S(x_1, x_3) \lor S(x_2, x_4)\big )\Big ] \land \Big [(\forall x_1)(\forall x_2)(\forall x_3)(\forall x_4)\\ & \big (E(x_1, x_2) \land E(x_2, x_3) \land E(x_3, x_4) \land E(x_1, x_4) \land \neg E(x_1, x_3) \land \neg E(x_2, x_4) \rightarrow S(x_1, x_2) \lor S(x_2, x_3) \lor S(x_3, x_4) \\ & \lor S(x_1, x_4)\big )\Big ] \land \Big [(\forall x_1)(\forall x_2)(\forall x_3)(\forall x_4)\big (E(x_1, x_2) \land E(x_2, x_3) \land E(x_3, x_4) \land \neg E(x_1, x_3) \land \neg E(x_1, x_4) \land \\ & \neg E(x_2, x_4) \rightarrow S(x_1, x_2) \lor S(x_2, x_3) \lor S(x_3, x_4)\big )\Big ]\Big \}.
\end{align*}
which ensures that whenever there is an induced $2K_2, C_4$ or $P_4$ on four vertices $x_1 ,x_2, x_3$ and $x_4$, atleast one of those edges is included in the solution set $S$.\\

\noindent MIN COMPARABILITY NODE DELETION: \\
Input:  A graph $G = (V, E)$.\\
Output:  MIN $S\subseteq V(G) \ni G\setminus S$ is a comparability graph.\\[0.3cm] 
The optimum solution for the problem can be expressed as follows:
\begin{align*}
& opt _{MIN\_COMP\_NODE\_DEL}(G) = min _{S}(O)\Big\{|S|:(G, S, O) \models \Big [ (\forall x)(\forall y)\big (E(x,y) \rightarrow \big (O(x,y) \land \\ & \neg O(y,x)\big ) \lor \big (O(y,x) \land \neg O(x,y)\big )\big ) \Big ] \land \Big [ (\forall x)(\forall y)(\forall z)\big (O(x,y) \land O(y,z) \land \neg O(x,z)\rightarrow S(x) \lor S(y) \lor S(z)\big )\Big ]\Big\}.
\end{align*}
where $O(x,y)$ implies orientation $\overrightarrow{(x,y)}$ and the expression ensures that either $O(x,y)$ or $O(y,x)$ is allowed but not both and whenever $O(x,y)$ and $O(y,z)$ are present and $O(x,z)$ is not present, atleast one of the three vertices $x,y$ and $z$ is present in the solution set $S$.\\

\noindent MIN COMPARABILITY EDGE DELETION: \\
Input:  A graph $G = (V, E)$.\\
Output:  MIN $S\subseteq E(G) \ni G=(V,E\setminus S)$ is a comparability graph.\\[0.3cm] 
The optimum solution for the problem can be expressed as follows:
\begin{align*}
& opt _{MIN\_COMP\_EDGE\_DEL}(G) = min _{S}(O)\Big \{|S|:(G, S, O) \models \Big [ (\forall x)(\forall y)\big (E(x,y) \land \big (O(x,y) \land \\ & \neg O(y,x)\big ) \lor \big (O(y,x) \land \neg O(x,y)\big )\big ) \Big ] \land \Big [ (\forall x)(\forall y)(\forall z)\big (O(x,y) \land O(y,z) \land \neg O(x,z) \rightarrow S(x,y) \lor S(y,z)\big )\Big ]\Big \}.
\end{align*}
where $O(x,y)$ implies orientation $\overrightarrow{(x,y)}$ and the expression ensures that either $O(x,y)$ or $O(y,x)$ is allowed but not both and whenever $O(x,y)$ and $O(y,z)$ are present and $O(x,z)$ is not present, atleast one of the two edges $\{x,y\}$ and $\{y,z\}$ is present in the solution set $S$.\\

\noindent MIN INTERVAL NODE DELETION: \\
Input:  A graph $G = (V, E)$.\\
Output:  MIN $S\subseteq V(G) \ni G\setminus S$ is a interval graph.\\[0.3cm] 
The optimum solution for the problem can be expressed as follows:
\begin{align*}
& opt _{MIN\_INT\_NODE\_DEL}(G) = min _{S}(O)\Big \{|S|:(G, S, O) \models \Big [ (\forall x)(\forall y)(\forall z)(\forall w)\big (E(x,y) \land E(y,z) \land E(z,w) \\ & \land E(w,x) \land \neg E(x,z) \land \neg E(y,w) \rightarrow S(x) \lor S(y) \lor S(z) \lor S(w)\big ) \Big ] \land \Big [ (\forall x)(\forall y)\big (\neg E(x,y) \rightarrow \big (O(x,y) \land \neg O(y,x)\big ) \\ & \lor \big (O(y,x) \land \neg O(x,y)\big )\big ) \Big ] \land \Big [ (\forall x)(\forall y)(\forall z)\big (O(x,y) \land O(y,z) \land \neg O(x,z) \rightarrow S(x) \lor S(y) \lor S(z)\big ) \Big ]\Big \}.
\end{align*}
where $O(x,y)$ implies orientation $\overrightarrow{(x,y)}$ and the expression ensures that whenever there is a cycle on 4 vertices $x, y, z$ and $w$, atleast one of those vertices is included in the solution set. And for every edge in $\overline{G}$ either $O(x,y)$ or $O(y,x)$ is allowed but not both and whenever $O(x,y)$ and $O(y,z)$ are present and $O(x,z)$ is not present, atleast one of the three vertices $x,y$ and $z$ is present in the solution set $S$.\\

\noindent MIN PERMUTATION NODE DELETION: \\
Input:  A graph $G = (V, E)$.\\
Output:  MIN $S\subseteq V(G) \ni G\setminus S$ is a permutation graph.\\[0.3cm] 
The optimum solution for the problem can be expressed as follows:
\begin{align*}
& opt _{MIN\_PERM\_NODE\_DEL}(G) = min _{S}(O_1,O_2)\Big \{|S|:(G, S, O_1, O_2) \models \Big [ (\forall x)(\forall y)\big (E(x,y) \rightarrow \\ & \big (O_1(x,y) \land \neg O_1(y,x)\big ) \lor \big (O_1(y,x) \land \neg O_1(x,y)\big )\big ) \Big ] \land \Big [ (\forall x)(\forall y)(\forall z)\big (O_1(x,y) \land O_1(y,z) \land \neg O_1(x,z)\rightarrow S(x) \lor \\ & S(y) \lor S(z)\big )\Big ] \land \Big [ (\forall x)(\forall y)\big (\neg E(x,y) \rightarrow \big (O_2(x,y) \land \neg O_2(y,x)\big ) \lor \big (O_2(y,x) \land \neg O_2(x,y)\big )\big ) \Big ] \land \Big [ (\forall x)(\forall y)(\forall z)  \\ & \big (O_2(x,y) \land O_2(y,z) \land \neg O_2(x,z) \rightarrow S(x) \lor S(y) \lor S(z)\big ) \Big ]\Big \}.
\end{align*}
where $O_1(x,y)$ or $O_2(x,y)$ implies orientation $\overrightarrow{(x,y)}$ and $O_1$ and $O_2$ are orientations in $G$ and $\overline{G}$ respectively. And the expression ensures that either $O_1(x,y)$ or $O_1(y,x)$ is allowed but not both and whenever $O_1(x,y)$ and $O_1(y,z)$ are present and $O_1(x,z)$ is not present, atleast one of the three vertices $x,y$ and $z$ are present in the solution set $S$. And for every edge in $\overline{G}$ either $O_2(x,y)$ or $O_2(y,x)$ is allowed but not both and whenever $O_2(x,y)$ and $O_2(y,z)$ is present and $O_2(x,z)$ is not present, atleast one of the three vertices $x,y$ and $z$ is present in the solution set $S$.

First order logic presented in this paper can be viewed as a monadic second order logic if we allow quantification over sets. That is, if we can express the property $\Psi$ allowing quantification over sets then for node/edge deletion problems, the resultant graph satisfying $\Psi$ (obtained by deleting a subset of vertices (edges)) can be verifed in linear time on graphs with bounded tree width. For example, node/edge deletion on a connected graph $G$ whose resultant is chordal or chordal bipartite is linear-time solvable on graphs with bounded tree width.  

\section{Results on MSO}
\label{mso}
\subsubsection{MSO for variants of coloring:} \hfill \\

\noindent Let the set $C$ which denotes the set of colors, $T(x,c)$ denotes that vertex $x$ is colored with color $c\in C$ and $T(x,y,c)$ denotes that the edge $\{x,y\}$ is colored with color $c\in C$. To express the variants of coloring in MSO, we make use of the some of the following properties expressed in MSO.\\

\noindent Coloring all the vertices of a graph $G = (V,E)$ is expressed in MSO as follows:
\begin{align*}
Color\_vertices(C) = (\forall x)(\exists c)(C(c) \land T(x,c))
\end{align*}

\noindent which ensures that for every vertex $x$ there exists some color $c\in C$. 

\noindent Coloring all the edges of a graph $G = (V,E)$ is expressed in MSO as follows:
\begin{align*}
Color\_edges(C) = (\forall x)(\forall y)(\exists c)(E(x,y) \land C(c) \land T(x,y,c))
\end{align*}

\noindent which ensures that for every edge $\{x, y\}$ there exists some color $c\in C$.

For a graph $G = (V, E)$, \emph{proper vertex coloring} is defined as coloring all the vertices in $G$ such that no two adjacent vertices have same color. This property is expressed in MSO as follows.

\begin{align*}
Proper\_Vertex\_Coloring(C) = Color\_vertices(C) \land [(\forall x)(\forall y)(\forall c_1)(\forall c_2)(E(x, y) \land T(x,c_1) \land T(y, c_2) \rightarrow c_1\neq c_2)]
\end{align*}

\noindent which ensures that all the vertices of $G$ are colored with some color and two colors $c_1$ and $c_2$ coloring any two adjacent vertices $x$ and $y$ respectively are different.\\

\noindent STAR COLORING \cite{reed}: Proper vertex coloring of a graph $G = (V,E)$ such that every path of three in $G$ uses atleast three different colors. This property is expressed in MSO as follows.

\begin{align*}
& Star\_Coloring(C) = Proper\_Vertex\_Coloring(C) \land (\forall x)(\forall y)(\forall z)(\forall w)(\forall c_1)(\forall c_2)(\forall c_3)(\forall c_4)(E(x,y) \land E(y,z) \\ & \land E(z,w)  \land T(x,c_1) \land T(y,c_2) \land T(z,c_3) \land T(w,c_4) \rightarrow (c_1 \neq c_2 \neq c_3 \neq c_4) \lor (c_1 = c_2 \neq c_3 \neq c_4) \lor \\ & (c_1 \neq c_2 = c_3 \neq c_4)  \lor (c_1 \neq c_2 \neq c_3 = c_4) \lor (c_1 = c_3 \neq c_2 \neq c_4) \lor (c_2 = c_4 \neq c_1 \neq c_3) \lor (c_1 = c_4 \neq c_2 \neq c_3))
\end{align*}

\noindent which ensures that all the vertices of $G$ are proper colored with some color and for every path on four vertices $x, y, z, w$ colored with the colors $c_1, c_2, c_3, c_4$ respectively, either all four colors are distinct or three colors are distinct.

For a given graph $G$ and a value $k$ which indicates the cardinality of $C$, the set $C$ satisfies Min Star Coloring if it satisfies the following property
\begin{align*}
Min\_Star\_Coloring(C) = (|C|\leq k) \land Star\_Coloring(C)
\end{align*}

\noindent CD COLORING: For a graph $G = (V, E)$, a partition of $V(G)$ into $l$ independent sets $V_1, V_2, \ldots , V_{l}$ is called $l-cd-coloring$ of $G$ if there exists a vertex $u_{i}\in V(G)$ such that $u_{i}$ dominates $V_{i}$ in $G$ for $1\leq i\leq l$. 

\noindent We say $x$ dominates $A$ if either (i) $A = \{x\}$ or (ii) $x\notin A$ and $(\forall a\in A)(\{x, a\}\in E(G))$. This property is expressed in MSO as follows.
\begin{align*}
Dominate(x, A) = (x\in A \land |A|=1) \lor (x\notin A \land (\forall a\in A)(E(x,a)))
\end{align*}

\noindent CD coloring is expressed in MSO as follows:
\begin{align*}
& CD\_Coloring(l) = (\exists V_1, \ldots , \exists V_{l})[(\forall i)((\forall x)(\forall y)(V_{i}(x) \land V_{i}(y) \land x\neq y \land \neg E(x,y))) \land (\forall i)(\exists u_{i}\in V)\\ &(Dominate (u_{i}, V_{i}))]
\end{align*}

\noindent which ensures $V(G)$ is partitioned into $l$ independent sets $V_1, V_2, \ldots , V_{l}$ and for every $V_{i}$ there exists a vertex $u_{i}$ which dominates it where $1\leq i\leq l$.

For a given graph $G$ and a value $k$ which indicates the value of $l$, $k$ satisfies Min CD Coloring if it satisfies the following property
\begin{align*}
Min\_CD\_Coloring(C) = (l\leq k) \land CD\_Coloring(C)
\end{align*}

\noindent EDGE COLORING: Edge coloring of a graph $G = (V, E)$ such that no two adjacent edges have same color. This property is expressed in MSO as follows.
\begin{align*}
& Edge\_Coloring(C) = Color\_edges(C) \land (\forall x)(\forall y)(\forall z)(\forall c_1)(\forall c_2)(E(x, y) \land E(y, z) \land x\neq z \land T(x, y, c_1) \land \\ & T(y, z, c_2)) \rightarrow c_1 \neq c_2)
\end{align*}

\noindent which ensures that all the edges of $G$ are colored with some color and two colors $c_1$ and $c_2$ coloring any two adjacent edges $\{x, y\}$ and $\{y, z\}$ respectively are different.

For a given graph $G$ and a value $k$ which indicates the cardinality of $C$, the set $C$ satisfies Min Edge Coloring if it satisfies the following property
\begin{align*}
Min\_Edge\_Coloring(C) = (|C|\leq k) \land Edge\_Coloring(C)
\end{align*}

\noindent RAINBOW COLORING: Edge coloring of a graph $G = (V, E)$ such that there is a rainbow path(no color repeats on the path) between each pair of its vertices. This property is expressed in MSO as follows.
\begin{align*}
& Rainbow\_Coloring(C) = Color\_edges(C) \land (\forall x)(\forall y)[E(x,y) \lor ((\forall c_0, \ldots , \forall c_{k}) \exists D\subset V \land (x_1, \ldots , x_{k})\in D \land \\ & E(x,x_1) \land (\bigwedge_{1\leq i \leq k-1} E(x_{i}, x_{i+1})) \land E(x_{k},y) \land T(x, x_1, c_0) \land (\bigwedge_{1\leq i \leq k-1} T(x_{i}, x_{i+1}, c_{i})) \land T(x_{k}, y, c_{k}) \rightarrow c_0\neq c_1\neq \ldots \neq c_{k})]
\end{align*}

\noindent which ensures that all the edges of $G$ are colored with some color and all the pair of vertices $x, y$ of path length greater than 1 with $x, x_1, \ldots , x_{k}, y$ as path are colored with different colors $c_0, \ldots , c_{k}$ respectively.

For a given graph $G$ and a value $k$ which indicates the cardinality of $C$, the set $C$ satisfies Min Rainbow Coloring if it satisfies the following property
\begin{align*}
Min\_Rainbow\_Coloring(C) = (|C|\leq k) \land Rainbow\_Coloring(C)
\end{align*}

\noindent TOTAL COLORING: Coloring of vertices and edges in a graph $G = (V, E)$ such that no adjacent edges and no edge and its end vertices are assigned the same color. This property is expressed in MSO as follows.
\begin{align*}
& Total\_Coloring(C) = Color\_vertices(C) \land Edge\_Coloring(C) \land [(\forall x)(\forall y)(E(x,y) \land T_1(x, y, c_1) \land T_2(x,c_2) \land \\ & T_2(y, c_3) \rightarrow c_1\neq c_2\neq c_3]
\end{align*}

\noindent which ensures that all the vertices and edges are covered with some color and no two adjacent edges are colored with same color and three colors coloring any edge $\{x, y\}$ with $c_1$ and the vertices $x$ and $y$ with $c_1$ and $c_2$ are different.

For a given graph $G$ and a value $k$ which indicates the cardinality of $C$, the set $C$ satisfies Min Total Coloring if it satisfies the following property
\begin{align*}
Min\_Total\_Coloring(C) = (|C|\leq k) \land Total\_Coloring(C)
\end{align*}

\noindent EQUITABLE COLORING: Proper vertex coloring of a graph $G = (V,E)$ such that no two adjacent vertices have the same color, and the numbers of vertices in any two color classes differ by at most one. This property is expressed in MSO as follows.
\begin{align*}
& Equitable\_Coloring(C) = Proper\_Vertex\_Coloring(C) \land [(\forall x)(\forall y)(\forall c_1)(\forall c_2)(c_1\neq c_2 \land \\ &  abs(|T(x,c_1)|-|T(y,c_2)|)\leq 1)]
\end{align*}

\noindent which ensures that all the vertices are covered with some color and no two vertices are colored with some color and for any two colors $c_1$ and $c_2$, absolute value of difference between the cardinalities of number of vertices colored with $c_1$ and $c_2$ is atmost 1.

For a given graph $G$ and a value $k$ which indicates the cardinality of $C$, the set $C$ satisfies Min Equitable Coloring if it satisfies the following property
\begin{align*}
Min\_Equitable\_Coloring(C) = (|C|\leq k) \land Equitable\_Coloring(C)
\end{align*}

\subsubsection{MSO for variants of dominating set:} \hfill

\noindent Let $S\subseteq V$ denote the solution set. To express the variants of dominating set in MSO, we make use of the some of the following properties expressed in MSO.

\noindent \emph{Connectedness} of a set $S$ can be expressed in MSO as follows:
\begin{align*}
& Connected(S) = (\forall x)(\forall y)(S(x) \land S(y) \land E(x,y) \lor (\exists P\subset S \land (x_1,x_2,\ldots ,x_{k})\in P \land E(x,x_1) \land \\ & (\bigwedge_{1\leq i \leq k-1} E(x_{i},x_{i+1})) \land E(x_{k},y)))
\end{align*}

\noindent which ensures that for every two vertices $x$ and $y$ in $S$ there is an edge between them or there exists a set $P\subset S$ that forms a path between them.\\

\noindent \emph{Cyclicity} of a set $S$ can be expressed in MSO as follows:
\begin{align*}
Cycle(S) = (\forall x)(\exists y)(\exists z)\big (S(x) \land S(y) \land S(z) \land y \neq z \land E(y,x) \land E(z,x) \land (\forall w\in [C-(y,z)])(\neg E(x,w))
\end{align*}

\noindent which ensures for every vertex $v$ in S, there exists two vertices $y$ and $z$ such that both are adjacent to $v$ which implies $S$ forms a cycle.

\noindent Checking if a set $S$ forms a \emph{clique} can be expressed in MSO as follows:
\begin{align*}
Clique(S) = (\forall x)(\forall y)(S(x) \land S(y) \land x\neq y \land E(x,y))
\end{align*}

\noindent which ensures that for every two vertices $x$ and $y$ in $S$ there is an edge between them. \\

\noindent For a graph $G = (V,E)$, a \emph{dominating set} is a set $S\subseteq V$ such that $S$ is connected and for every vertex $x\in V$ is in $S$ or $x$ is adjacent to a vertex in $S$. It can be expressed in MSO as follows:
\begin{align*}
Dom(S) = (\forall x)(\exists y)(S(x) \lor (S(y) \land E(x,y)))
\end{align*}

\noindent CONNECTED DOMINATING SET: For a graph $G = (V,E)$, a set $S\subseteq V$ such that $S$ is connected and for every vertex $x\in V$ is in $S$ or $x$ is adjacent to a vertex in $S$. This property is expressed in MSO as follows:
\begin{align*}
Connected\_Dom(S) = Dom(S) \land Connected(S)
\end{align*}

For a given graph $G$ and a value $k$ which indicates the cardinality of $S$, the set $S$ satisfies Min Connected Dominating Set if it satisfies the following property
\begin{align*}
Min\_Connected\_Dom(S) = (|S|\leq k) \land Connected\_Dom(S)
\end{align*}

\noindent TOTAL DOMINATING SET:
For a graph $G = (V,E)$, a set $S\subseteq V$ such that every vertex $x\in V$ is adjacent to some vertex in $S$. This property is expressed in MSO as follows:
\begin{align*}
Total\_Dom(S) = (\forall x)(\exists y)(S(y) \land E(x,y))
\end{align*}

\noindent which ensures that there is some vertex $y$ in $S$ adjacent to $x$.\\

For a given graph $G$ and a value $k$ which indicates the cardinality of $S$, the set $S$ satisfies Min Total Dominating Set if it satisfies the following property
\begin{align*}
Min\_Total\_Dom(S) = (|S|\leq k) \land Total\_Dom(S)
\end{align*}

\noindent TOTAL OUTER CONNECTED DOMINATING SET:
For a graph $G = (V,E)$, a set $S\subseteq V$ such that for every vertex $x\in V$ is adjacent to some vertex in $S$ and the subgraph induced by $V\setminus S$ is connected. This property is expressed in MSO as follows:
\begin{align*}
Total\_Outer\_Connected\_Dom(S) = Total\_Dom(S) \land Connected(V\setminus S)
\end{align*}
where
\begin{align*}
& Connected(V\setminus S) = (\forall x)(\forall y)(\neg S(x) \land \neg S(y) \land \big (E(x,y) \lor \big (\exists P \subset V\setminus S \land (x_{1}, x_{2}, \ldots , x_{k})\in P \land E(x,x_{1}) \land \\ & (\bigwedge_{1\leq i \leq k-1} E(x_{i},x_{i+1})) \land E(x_{k},y))\big )\big )
\end{align*}

\noindent which that ensures for every two vertices not in $S$, there is an edge or there exists a set $P\subset S$ which forms a path between them.\\

For a given graph $G$ and a value $k$ which indicates the cardinality of $S$, the set $S$ satisfies Min Total Outer Connected Dominating Set if it satisfies the following property
\begin{align*}
Min\_Total\_Outer\_Connected\_Dom(S) = (|S|\leq k) \land Total\_Outer\_Connected\_Dom(S)
\end{align*}

\noindent CYCLE DOMINATING SET:
For a graph $G = (V,E)$, a set $S\subseteq V$ such that $S$ forms a cycle in $G$ and for every vertex $x\in V$ is in $S$ or $x$ is adjacent to a vertex in $S$. This property is expressed in MSO as follows:
\begin{align*}
Cycle\_Dom(S) = Dom(S) \land Cycle(S) \land connected(S)
\end{align*}

\noindent which ensures that $S$ is a dominating set and forms a cycle.\\

For a given graph $G$ and a value $k$ which indicates the cardinality of $S$, the set $S$ satisfies Min Cycle Dominating Set if it satisfies the following property
\begin{align*}
Min\_Cycle\_Dom(S) = (|S|\leq k) \land Cycle\_Dom(S)
\end{align*}

\noindent PERFECT DOMINATING SET:
For a graph $G = (V,E)$, a set $S\subseteq V$ such that $S$ every vertex $x\notin S$ is adjacent to exactly one vertex in $S$. This property is expressed in MSO as follows:
\begin{align*}
Perfect\_Dom(S) = (\forall x)(\exists y)(\forall z)(\neg S(x) \land S(y) \land E(x, y) \land S(z) \land E(x,z) \rightarrow y=z)
\end{align*}

\noindent which ensures that for every vertex $x\notin S$ there exists some vertex $y\in S$ to which $x$ is adjacent and if $x$ is adjacent to any other vertex $z\in S$, then $y$ is equal to $z$.\\

For a given graph $G$ and a value $k$ which indicates the cardinality of $S$, the set $S$ satisfies Min Perfect Dominating Set if it satisfies the following property
\begin{align*}
Min\_Perfect\_Dom(S) = (|S|\leq k) \land Perfect\_Dom(S)
\end{align*}

\noindent CLIQUE DOMINATING SET:
For a graph $G = (V,E)$, a set $S\subseteq V$ such that $S$ is a clique and for every vertex $x\in V$ is in $S$ or $x$ is adjacent to a vertex in $S$. This property is expressed in MSO as follows:
\begin{align*}
Clique\_Dom(S) = Dom(S) \land Clique(S)
\end{align*}

\noindent which ensures that $S$ is a dominating set and forms a clique.\\

For a given graph $G$ and a value $k$ which indicates the cardinality of $S$, the set $S$ satisfies Min Perfect Dominating Set if it satisfies the following property
\begin{align*}
Min\_Clique\_Dom(S) = (|S|\leq k) \land Clique\_Dom(S)
\end{align*}

\section{Directions for further research}         
In this paper, we have presented FO and MSO for various graph-theoretic problems. As a consequence, it follows that we have constant-factor approximation algorithms or linear-time algorithms on bounded tree-width graphs. As these results are existential in nature, an interesting direction would be to develop algorithms with good approximation ratios.


\begin{thebibliography}{4}

\bibitem{andras}
A. Brandst{\"a}dt, Van Bang Le and J. P. Spinrad, Graph classes: a survey, 1999, SIAM.

\bibitem{courcelle}
B. Courcelle, M. Mosbah, Monadic second-order evaluations on tree-decomposable graphs, Theoretical Computer Science, 109 (1), 1993, 49-82.

\bibitem{domappln}
Du. Ding-Zhu, Wan.Peng-Jun, Connected Dominating Set: Theory and Applications, Springer Optimization and Its Applications, 77, 2013.

\bibitem{fagin}
R. Fagin,  Generalized first order spectra and polynomial time recognizable sets. Complexity of Computations, SIAM-AMS Proc. 7, 1974, 43-73.

\bibitem{reed}
G. Fertin, A. Rauspad, B. Reed, Star coloring of graphs, Journal of Graph theory, 2004.

\bibitem{unifiedframework}
T. Fujito, A unified approximation algorithm problems, Discrete Applied Mathematics, 86, (1998), 213-23.

\bibitem{golumbic}
M. C. Golumbic, Algorithmic graph theory and perfect graphs, volume 57. Elsevier, 2004.

\bibitem{kolaitis}
P. G. Kolaitis, M. N. Thakur, Approximation properties of NP minimization classes, Journal of Computer and System Sciences, 50(3), 1995, 391-411.


\bibitem{appln}
J. M. Lewis, M. Yannakakis, The Node-deletion problem for hereditary properties is NP-complete,
J. Comput. System Sci. 20 (1980) 219-230.

\bibitem{maxsubgraph}
C. Lund, M. Yannakakis, The approximation of maximum subgraph problems, in: Proc. 20th ICALP, Lecture Notes in Computer Science, vol. 700, Springer, 1993, pp. 40-51.

\bibitem{papad}
C. H. Papadimitriou, M. Yannakakis, Optimization, approximation, and complexity classes, Journal of computer and system sciences, 43(3), 1991, 425-440.

\bibitem{west}
D. B. West, Introduction to graph theory, volume 2. Prentice hall Upper Saddle River, 2001.

\end{thebibliography}
\end{document}